# A PATH TO HIGHER $Q_0$ WITH LARGE GRAIN NIOBIUM CAVITIES*

Pashupati Dhakal[#], Gianluigi Ciovati, and Ganapati Rao Myneni
Jefferson Lab, Newport News VA 23606, USA

*Abstract*

The improvement of the quality factor $Q_0$ of superconducting radio-frequency (SRF) cavities at medium accelerating gradients (20-25 MV/m) is important in order to reduce the cryogenic losses in continuous wave (CW) accelerators used for a variety of applications. In recent years, SRF cavities fabricated from ingot niobium have become a viable alternative to standard high-purity fine-grain Nb for the fabrication of high-performing SRF cavities with the possibility of significant cost reduction. Recent studies demonstrated the improvement of $Q_0$ at medium field in cavities heat treated at 800–1200 °C without subsequent chemical etching [1]. To further explore this treatment procedure, a new induction furnace with an all-niobium hot-zone was commissioned [2]. A single-cell 1.5 GHz cavity fabricated from ingot material from CBMM, Brazil, with RRR~200, was heat treated in the new furnace in the temperature range 800–1400 °C for several hours. Residual resistance value of $1 - 5$ n$\Omega$ have been consistently achieved on this cavity $Q_0$-values as high as $4.6\times10^{10}$ at 90 mT peak surface magnetic field at 2 K. $Q_0$ values of about $\sim 2\times10^{11}$ have been measured at 1.5 K.

## INTRODUCTION

The overall performance of SRF cavities is measured by the dependence of Q0-values on accelerating gradient, where $Q_0=G/R_s$ with $G$ is the geometric factor which depends on the cavity geometry and $R_s$ is the surface resistance. Higher $Q_0$ for the reduction of cryogenic loss in CW operation and higher $E_{acc}$ for high energy accelerators are desired. In the last four decades, much work has been done to push the accelerating gradient to higher and higher values. The increase in quality factor is equally important to reduce the operating cost of the future CW accelerators. In this contribution, we present the recent results on single cell cavities subjected to the several combinations of high-temperature heat-treatments (HT) and low-temperature baking (LTB) to reduce the surface resistance and hence increase the quality factor.

The surface resistance in superconducting materials is the sum of the temperature dependent BCS resistance and temperature independent residual resistance. The sources of the $R_{res}$ are the trapped magnetic flux during the cavity cool down, impurities, hydrides and oxides, imperfections, and surface contamination. The BCS surface resistance is calculated using the BCS theory of superconductivity approximated for the temperature range $T<T_c/2$ as

$$R_{BCS}(T,f) = (Af^2/T)\, e^{-\Delta/k_BT} \qquad (1)$$

where $f$ is the resonant frequency, $\Delta$ is the superconducting gap at 0 K, $k_B$ is the Boltzmann constant and $A$ is a factor which depends on material parameters such as the coherence length ($\xi$), the London penetration depth ($\lambda_L$) and the electrons mean free path ($l$).

Material parameters such as the transition temperature, superconducting gap, coherence length and London penetration depth are not expected to change significantly with different cavity preparation procedures. However, the electronic mean free path can be easily affected by the purity of the starting Nb material as well as the cavity treatment procedures. The values of $R_{BCS}$ calculated from the full BCS theory as a function of electronic mean free path for Nb at 1.5 GHz at different temperature, using the computer code developed by Halbritter [3], is shown in Fig. 1. We have used the material parameters, $\Delta/k_BT = 1.82$, $\lambda_L = 32$ nm, $\xi = 39$ nm, $T_c = 9.25$ K, which describe Nb after the typical cavity surface preparation process [4]. The BCS resistance has a minimum when the electronic mean free path is of the order of the coherence length. Figure 1 shows that a Nb cavity of lower purity near the RF surface could have a factor of ~2 lower $R_{BCS}$ than one with higher residual resistivity ratio at the surface.

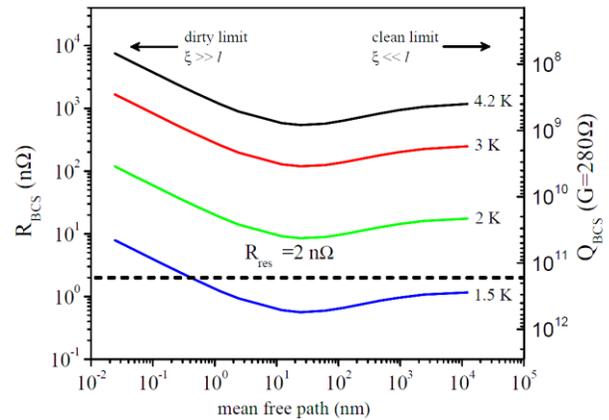

Figure 1: The BCS resistance as a function of electron mean free path at frequency 1.5 GHz. The minimum in $R_{BCS}$ is for a mean free path value which corresponds to a residual resistivity ratio (RRR) of about 13. The dotted line with $R_{res} = 2$ n$\Omega$ is drawn for illustrative purpose for residual resistance. The total surface resistance is the sum of the $R_{res}$ and $R_{BCS}$. Also shown is the corresponding $Q_{BCS}$ for an elliptical cavity with geometry factor $G = 280\, \Omega$.

___

## INGOT NIOBIUM

The current state-of-art SRF technology is relying on high-purity (residual resistivity ratio > 300), fine-grain (ASTM > 5) bulk niobium. In recent years, large grain (LG) ingot Nb became an alternative to the fine grain Nb for the fabrication of high performance SRF cavities. Simpler fabrication procedures, potential cost reduction, higher thermal stability at 2K due to the enhanced thermal conductivity [5], as well as reproducibility in the performance of cavities has attracted the SRF community towards the fabrication of SRF cavities with ingot Nb [6,7]. Multi-cell cavities made of LG ingot have achieved accelerating gradients as high as the fine grain counterparts [8]. A recent study suggested that LG cavities could have lower $R_{res}$ due to trapped magnetic flux because of lower pinning centres, compared to fine-grain Nb [9], giving rise to higher $Q_0$ values.

## CAVITY PREPERATION AND TEST RESULTS

A 1.47 GHz single cell cavity, original CEBAF shape, was fabricated from ingot−G (~1375 wt.ppm Ta content, RRR~200) from CBMM, Brazil. The cavity was subjected to 73 μm material removal by centrifugal barrel polishing and 65 μm buffered chemical polishing (BCP 1:1:2). Subsequently, the cavity was heat treated in the new induction furnace at different temperatures as listed below. After all the heat treatment processes, the cavity was cooled down in the presence of Argon gas with partial pressure of ~ $10^{-6}$ torr as a way to purge residual gases in the furnace. Once the cavity is cooled down to room temperature, the furnace is vented with high purity Oxygen for a controlled oxidation of the surface. The treatment and test sequence was as follows:

- ~20 μm removal by BCP 1:1:2
- High-power RF test at 2.0 K (baseline 1)
- HT at 800 °C/6 h
- High-power RF test at 2.0 K
- In-situ LTB 120 °C/12 h
- High-power RF test at 2.0 K
- 15 μm removal by BCP 1:1:2
- High-power RF test at 2.0 K (baseline 2)
- HT at 1000 °C/6 h
- High-power RF test at 2.0 K
- In-situ LTB 120 °C/12 h
- High-power RF test at 2.0 K
- 29 μm removal by BCP 1:1:2
- High-power RF test at 2.0 K (baseline 3)
- HT at 1200 °C/6 h
- High-power RF test at 2.0 K
- In-situ LTB 120 °C/12 h
- High-power RF test at 2.0 K
- 29 μm removal by BCP 1:1:2
- High-power RF test at 2.0 K (baseline 4)
- HT at 1400 °C/3 h
- High-power RF test at 2.0 K
- In-situ LTB 120 °C/12 h
- High-power RF test at 2.0 K

Common to the preparation steps mentioned above are: ultrasonic degreasing for 30 min prior to BCP or HT and after HT, High Pressure Rinse (HPR) with DI water for 1 h, after BCP or HT. After HPR the cavity is dried for ~3 h in a class 10 clean room and stainless steel blanks with pump-out port and RF antenna are assembled on the cavity flanges with indium wire as gasket. The cavity is then evacuated to ~$10^{-8}$ mbar on a vertical stand and the thermometry system is attached [10]. The RF tests consisted of measurements of $Q_0$ ($T$, ~10 mT) during the pump-down of the helium bath and $Q_0$ (2 K, $B_p$) and. Temperature maps of the outer cavity surface were also taken during the RF tests.

Table 1
Summary of the RF test results

| Test | $R_{res}$ (nΩ) | $\Delta/k_BT_c$ | $Q_0$(2 K,90 mT) |
|---|---|---|---|
| Baseline 1 | < 1.0 | 1.70±0.02 | $1.6 \times 10^{10}$ |
| 800 °C/6 h | 3.2 ± 0.2 | 1.81±0.02 | $2.0 \times 10^{10}$ |
| 120 °C/24 h | 6.3 ± 0.5 | 1.84±0.01 | $1.8 \times 10^{10}$ |
| Baseline 2 | 1.0 ± 1.0 | 1.80±0.03 | $1.9 \times 10^{10}$ |
| 1000 °C/6 h | 1.4 ± 0.4 | 1.86±0.02 | $2.2 \times 10^{10}$ |
| 120 °C/12 h | 2.0 ± 0.2 | 1.86±0.02 | $2.1 \times 10^{10}$ |
| Baseline 3 | 2.7 ± 0.3 | 1.81±0.02 | $1.8 \times 10^{10}$ |
| 1200 °C/6 h | <1.0 | 1.86±0.02 | $2.2 \times 10^{10}$ |
| 120 °C/12 h | 1.6 ± 0.3 | 1.86±0.02 | $2.3 \times 10^{10}$ |
| Baseline 4 | 2.0 ± 0.2 | 1.87±0.02 | $2.2 \times 10^{10}$ |
| 1400 °C/3 h | 1.0 ± 0.2 | 1.90±0.01 | $4.6 \times 10^{10}$ |
| 120 °C/12 h | 2.8 ± 0.8 | 1.93±0.02 | $4.0 \times 10^{10}$ |

Table 1 shows the summary of the test results of the cavity subjected to several heat treatments. Figure 2a shows the surface resistance measured at $B_p \cong 10$ mT in the temperature range of 4.3–1.5 K for baseline 4 and after HT at 1400 °C. The data are fitted using the BCS theory to extract $R_{res}$, mean free path and $\Delta/k_BT_c$. Figure 2 b shows the results of the $Q_0$ vs $B_p$ before and after the 1400 °C heat treatment. The baseline RF measurements showed a $Q_0$ of $2.2\times10^{10}$ at low field which stayed nearly constant until it quenched at ~100 mT. However, after the cavity is subjected to the HT at 1400 °C, the low field $Q_0$ (2K, 10mT) was $2.85\times10^{10}$ and increased up to $4.6\times10^{10}$ before it quenched at ~90 mT. This value of $Q_0$ is the highest ever measured in an SRF niobium cavity at 1.5 GHz and 2 K. The low-field $Q_0$ at 1.5 K measured after HT at 1400 °C was ~$2\times10^{11}$. No field emission was detected in any of the tests. The quench location (Fig. 3) was observed near the equator where a grain boundary was present and did not change after HT. Although a so-called low-field Q-increase is sometimes observed in SRF Nb cavities, it usually saturates at ~20–30 mT [3]. After HT at 1400 °C, however, this effect persisted up to ~60 mT. The origin of the low-field Q-increase in SRF Nb cavities is still unknown and under further study. The test results after HT shown in Fig. 2b were reproduced when using a different RF system and operator. In addition, it was checked that there was no significant power dependence of the insertion loss of the high-power RF cable inside the cryostat. The cavity performance after LTB showed the same behavior as that after HT at 1400 °C, but due to the increase in $R_{res}$ the $Q_0$–values is decreased by ~10%.

## CONCLUSIONS

We have measured the highest ever quality factor in 1.5 GHz SRF cavities on a single-cell made of large grain ingot niobium. The cavity was heat treated at 1400 °C/3 h in a newly commissioned all niobium hot-zone UHV induction furnace, with no subsequent chemical etching.

Small samples which were heat treated with the cavity are under study to characterize the surface properties after heat treatment. The study includes the SIMS, TEM and point contact tunneling experiments and the results will be presented in a future publication.

The successful development of such heat treatment could result in large-grain Nb being the material of choice for SRF cavities for future accelerators and less expensive cavity preparation procedure.

## ACKNOWLEDGEMENTS

We also would like to acknowledge T. Harris and J. Davenport for helping with the HPR and BCP and P. Kushnick for cryogenic support.

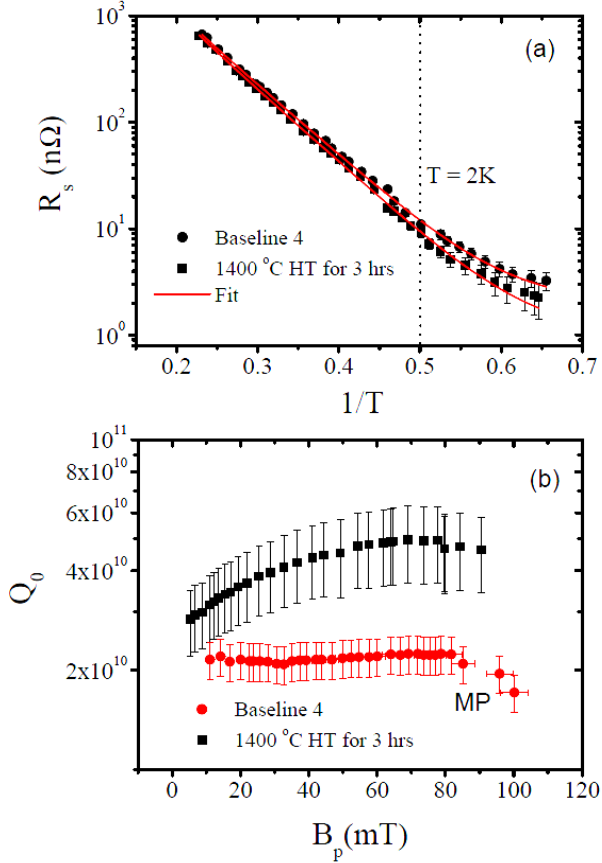

Figure 2: (a) Surface resistance $R_s$ as a function of temperature before and after 1400 °C heat treatment. (b) $Q_0$ ($B_p$) measured at 2.0 K. The tests were limited by quench.

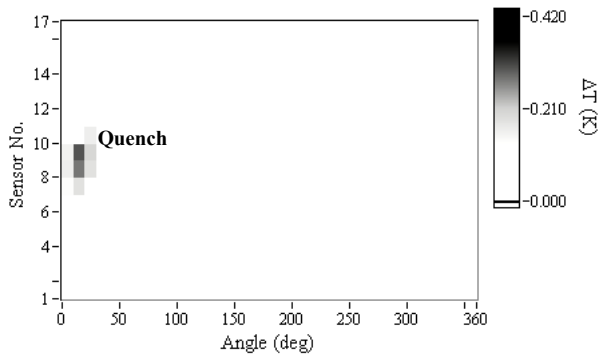

Figure 3: "Unfolded" temperature map at 2.0 K during quench at $B_p$ = 90 mT, after 1400 °C HT. The equator weld is between sensor No. 8 and 9.